\documentclass[a4paper,pra,twocolumn,superscriptaddress]{revtex4}
\usepackage{graphicx}
\usepackage{times}
\usepackage{amsmath}
\usepackage{hyperref}

\begin{document}

\title{Maximizing the electric field strength in the foci of high
  numerical aperture optics}

\author{Markus Sondermann}
\email{markus.sondermann@physik.uni-erlangen.de}
\affiliation{Institute of Optics, Information and Photonics,
  University of Erlangen-Nuremberg, 91058 Erlangen, Germany}
\affiliation{Max Planck Institute for the Science of Light, 91058
  Erlangen, Germany}
\author{Norbert Lindlein}
\affiliation{Institute of Optics, Information and Photonics,
  University of Erlangen-Nuremberg, 91058 Erlangen, Germany}
\author{Gerd Leuchs}
\affiliation{Institute of Optics, Information and Photonics,
  University of Erlangen-Nuremberg, 91058 Erlangen, Germany}
\affiliation{Max Planck Institute for the Science of Light, 91058
  Erlangen, Germany}

\begin{abstract}
Several applications require spatial distributions of the incident
electric field that maximize the electric field at a focal point for
a given input power.
The field distributions are derived for various optical systems in a
direct way based on fundamental physical properties.
The results may prove useful for a wide range of applications, e.g.,
microscopy, scattering experiments or excitation of single atoms.
For commonly used distributions - fundamental Gaussian modes and
doughnut modes - we give the upper bounds of the achievable field
amplitudes.
\end{abstract}

\maketitle

\section{Introduction}

Recently, there has been increased interest in the interaction of
single matter systems like atoms, molecules or ions with weak light
beams or single photons in free space.
The topical spread covers scattering by single two-level systems
\cite{vamivakas2007,wrigge2008,tey2008-np,zumofen2008,tey2009}
(including electromagnetically induced
transparency\cite{slodicka2010}) as well as the absorption of single
photons by single
atoms~\cite{lindlein2007,sondermann2007,pinotsi2008,piro2011}.
Also other applications relying on high numerical aperture (NA) optics
have raised increasing attention, involving parabolic
mirrors~\cite{stadler2008,bokor2008} as well as
lenses~\cite{bokor2004,urbach2008}. 

It is common to all these applications that they require the smallest
possible focal volume or -- in other words -- that the power incident
onto the focusing optics is transformed in such a way that the
electric energy density in the focus is maximized.
Thus, one wants to maximize the amplitude of the electric field in the
focus. 
Although it is well known since quite some time that the optimum
performance is achieved by electric dipole radiation
\cite{basset1986,stamnes1996,sheppard1997,quabis2000} and also some
newer publications hint in this direction \cite{bokor2004,vanenk2004}, this
recipe is followed only in few papers in a consequent manner
\cite{lindlein2007,sondermann2007,zumofen2008}.
There are also publications in which some sort of 'optimization
procedure' is followed: 
In Ref. \cite{tey2009} the beam width of a circularly polarized
fundamental Gaussian is varied in order to maximize the strength 
of the circularly polarized field component in the focus of an
aspheric lens. 
The authors of Ref. \cite{bokor2008} follow a similar procedure in
varying the beam waist of a radially polarized doughnut beam during
minimization of the focal volume in the case of a parabolic mirror.
In both cases, the optimization procedure is equivalent to maximizing
the overlap of the incident radiation with dipole radiation.
In the former case \cite{tey2009} the Gaussian fundamental mode is
matched towards the distribution that creates the field of a circular
dipole after refraction by the focusing lens.
In the latter case \cite{bokor2008} the doughnut mode is adapted such
that it best overlaps with the field of a linear dipole that has its
axis parallel to the optical axis of the parabolic mirror.
Finally, in a recent paper the field distribution that maximizes the
focal field component parallel to the optical axis of a high NA lens
is directly calculated by a variational method \cite{urbach2008} (this
concept has been extended to arbitrary orientation of the field vector
in Ref. \cite{urbach2009}).
However, with the exemption of
Refs.~\cite{bokor2008,bokor2004,chen2010b} no direct and explicit
relation between the found optimal radiation modes at the input and a
dipole radiation pattern is given in these publications.

Based on earlier results obtained by others
\cite{basset1986,sheppard1997} we follow a different approach in what
follows here.  
Knowing that it is only electric dipole radiation that contributes to
the focal electric field (see, e.g.,
Refs. \cite{basset1986,vanenk2004}), we calculate the distribution of
the field incident onto the focusing device that gives the desired
dipole radiation pattern after reflection/refraction by the focusing
optics, as it was done for a parabolic mirror and linear dipole
radiation in Ref. \cite{lindlein2007}.
This approach is on the one hand the most direct one.
On the other hand, it is conceptually simple in comparison to other
methods.
Besides temporal/spectral issues not treated here, this ansatz is the
same as the time reversal argument given in
Refs. \cite{lindlein2007,sondermann2007,quabis2000} (see also, e.g.,
Ref. \cite{lerosey2004} for time reversal focusing of microwaves
and Ref. \cite{derosny2002} for acoustic waves). 
\footnote{We note that recently the approach we follow here has been
  applied for the case of an aplanatic lens \cite{chen2010b}.}

Some of the results obtained here can in parts also be found in
other publications
\cite{sheppard1997,stamnes1996,sheppard1997-a,dhayalan1997}. 
However, a comprehensive study that covers all practically relevant
cases is still lacking.
Therefore, this paper is intended to deliver a recipe for maximizing
the electric field in focusing applications, in particular the
excitation of single atoms as envisaged in
Refs. \cite{sondermann2007,tey2009,tey2008-np,zumofen2008,piro2011}.
In the next section, we will emphasize the importance of a large solid
angle of illumination. 
In Secs. \ref{sec:intensity} and \ref{sec:polarization} the optimal
intensity and polarization distributions will be derived for the cases
of illumination by means of a parabolic mirror, a lens fulfilling the
sine condition and an ideal thin lens.
Since the experimental realization of these distributions may turn out
to be rather involved, the overlap of 'standard' radiation modes with
the ideal ones is treated in Sec. \ref{sec:overlap}.

Unlike in many other publications where the field in the focal region
is treated, the discussion will be restricted to the electric field in
the very focal point itself.
For the main applications in mind here -- scattering by single atoms or
absorption of single photons -- this constitutes a well justified
restriction, since it is only the field at the location of the atom
that matters (see also Ref. \cite{vanenk2004,tey2009,zumofen2008}).
Furthermore, this restriction leads to a simple analytical
result for the field in the focal point.

\section{Influence of the solid angle}
\label{sec:solid_angle}

Before starting with the calculations, we make some notes on the
terminology used in this paper.
For the quantum-optical applications mentioned above, the field
parallel to an atomic dipole has to be maximized.
In these cases, the 'dipole axis' is given by the quantization axis of
the scenario at hand.
In a classical sense, the term 'dipole axis' designates the direction
from which the polar angle $\vartheta$ is measured that
parametrizes the angular dipole radiation patterns
(see also Fig. \ref{fig:coords}). 
However, one may also think of applications in which no particular
emitter or source is present at the focus but one wants to maximize
the electric field vector that points into a certain direction (see
also Ref.~\cite{urbach2009}).
This field vector can be thought of as being parallel to some
imagined dipole moment. 
Thus, we will adopt the terminology of Ref. \cite{zumofen2008} and
often speak of 'virtual dipoles', i.e., dipoles that \emph{would} 
produce a certain kind of radiation pattern but are not present in
reality.

The term 'solid angle' usually designates the integral over the covered
polar angle and the covered azimuthal angle. 
Here, we will use the term 'weighted solid angle'.
It designates the covered solid angle weighted by a certain dipole
emission pattern:
$\Omega=\int\!{D}(\vartheta)\sin{\vartheta}d\vartheta d\varphi$,
with $D_\pi(\vartheta)=\sin^2{\vartheta}$ for a linear dipole or
$D_{\sigma_\pm}(\vartheta)=(1+\cos^2{\vartheta})/2$ for a circular
dipole~\cite{jackson1999}, respectively.
Thus, the maximum achievable, weighted solid angle is
$\Omega_\textrm{max}=8\pi/3$ instead of $4\pi$ in the usual
sense. 
For an incomplete solid angle parametrized by the limits
$\vartheta_{1,2}$, $\varphi_{1,2}$ one has  
\begin{equation}
\Omega_\pi=
\left[\frac{\cos^3\vartheta}{3}-\cos\vartheta
\right]_{\vartheta_1}^{\vartheta_2}
\cdot (\varphi_2-\varphi_1)
\end{equation}
for a linear dipole and 
\begin{equation}
\Omega_{\sigma_\pm}=\frac{1}{2}
\left[-\frac{\cos^3\vartheta}{3}-\cos\vartheta
\right]_{\vartheta_1}^{\vartheta_2}
\cdot (\varphi_2-\varphi_1)
\end{equation}
for a circular dipole.

We now derive an analytic expression that highlights the influence of
the covered weighted solid angle.
The electro-magnetic field can be decomposed into its different
multipole components.
This has been exploited in Ref. \cite{sheppard1997-a} to calculate the
field in the focal region for a given input field.
However, since we are only interested in the field at the focus, which we
take to coincide with the origin of our coordinate system, we only
need to take care of the dipole components.
Furthermore, it has already been pointed out in
Ref. \cite{stamnes1996} that each kind of dipole radiation creates
predominantly its respective state of polarization at the
focus/origin.
Therefore, it is obvious that one only needs to create the radiation
pattern of a single kind of dipole for most practical cases.

In order to calculate the (time independent) field in the focus, we
use the Debye-integral in the form given in Ref. \cite{lieb2001}:
\begin{equation}
\label{eq:debye}
\mathbf{E}(0)= -i\frac{f}{\lambda} 
\int \mathbf{E}_{f}(\vartheta,\varphi)\sin{\vartheta}\ d\vartheta d
\varphi\quad ,
\end{equation}
where $\mathbf{E}_{f}(\vartheta,\varphi)$ is the field on the focal
sphere for a focal length $f$, $\varphi$ is the azimuthal angle and
$\lambda$ the wavelength of the light to be focused.
Furthermore, a term $\exp(i\mathbf{k}\cdot\mathbf{r})$ has already been dropped
since we treat only the case $\mathbf{r}=0$.

If only a single kind of dipole radiation is incident onto the focus,
the field on the focal sphere can be written as
$\mathbf{E}_f=\psi_d\mathbf{p}_d$, where $\psi_d$ is a proportionality factor and
$\mathbf{p}_d$ is the polarization vector of dipole radiation in the
far field with $d=\pi,\sigma_\pm$ for a linear or circular dipole,
respectively. 
The polarization vectors in spherical coordinates are given by
\cite{wangsness1986}
\begin{equation}
\label{eq:polvec_lin}
\mathbf{p}_\pi=-\sin\vartheta\mathbf{e}_\vartheta
\end{equation}
 and 
\begin{equation}
\label{eq:polvec_circ}
\mathbf{p}_{\sigma_\pm}=\frac{1}{\sqrt{2}}
(\cos\vartheta[\cos\varphi \pm i\sin\varphi]\mathbf{e}_\vartheta 
+ [\pm i\cos\varphi - \sin\varphi]\mathbf{e}_\varphi)
\end{equation}
with $\mathbf{e}_{\vartheta,\varphi}$ being the polar and azimuthal
unit vectors. 
Inserting the corresponding polarization vector $\mathbf{p}_d$ 
and integrating over the full solid angle for a linear dipole as well as for
circular dipoles, the focal field strength is given by 
\begin{equation}
E_d^{\textrm{max}}(0)=-i\frac{f}{\lambda}\psi_d\cdot\frac{8\pi}{3}
\end{equation}
with the field vector parallel to the (virtual) dipole moment in the focus.

The total power radiated through the surface of the focal sphere 
can be written as
\begin{equation}
P=\frac{1}{2}\epsilon_0 c f^2\int\limits_{0}^{2\pi} 
\int\limits_{0}^{\pi} 
\psi_d^2 |\mathbf{p}_d|^2 \sin\vartheta\ d\vartheta d\varphi
\quad .
\end{equation}
The above expression is solved for $\psi_d$:
\begin{equation}
\psi_d=\frac{\sqrt{2P}}{f\sqrt{\epsilon_0c}}\cdot\sqrt{\frac{3}{8\pi}} \quad.
\end{equation}
This results in the maximum possible field amplitude 
\begin{equation}
\label{eq:maxfield}
E_d^{\textrm{max}}(0)=-i\frac{\sqrt{2P}}{\lambda\sqrt{\epsilon_0c}} \cdot
\sqrt{\frac{8\pi}{3}} \quad .
\end{equation}
This field delivers the maximum possible electric energy density given
earlier by Bassett \cite{basset1986}.

We now generalize the treatment to incident fields that do not
constitute a perfect dipole wave.
In this case only a fraction of the field given by
Eq.~\ref{eq:maxfield} is expected to be created at the focus.
Expanding the incident field on the focal sphere in multipoles and
projecting onto the dipole field of interest gives
\begin{equation}
\label{eq:field_gamma}
E_d^{\textrm{max}}(0)=-i\frac{\sqrt{2P}}{\lambda\sqrt{\epsilon_0c}} \cdot
\sqrt{\frac{8\pi}{3}} \cdot\gamma
\end{equation}
with 
\begin{equation}
\label{eq:gamma}
\gamma=\frac{
\int\limits_{0}^{2\pi} \int\limits_{0}^{\pi}\mathbf{p}_d\cdot\mathbf{E}_{f}
\sin\vartheta\ d\vartheta d\varphi
}
{
\sqrt{\int\limits_{0}^{2\pi} \int\limits_{0}^{\pi}|\mathbf{p}_d|^2
\sin\vartheta\ d\vartheta d\varphi}
\cdot
\sqrt{\int\limits_{0}^{2\pi} \int\limits_{0}^{\pi}|\mathbf{E}_f|^2
\sin\vartheta\ d\vartheta d\varphi}
}\quad .
\end{equation}
All integrals involving $\mathbf{E}_f$ in the definition of $\gamma$ need only to be
taken in the limits $\vartheta_{1,2}$, $\varphi_{1,2}$, i.e., on the
part of the solid angle for which $\mathbf{E}_f\ne0$.
Expanding Eq.~\ref{eq:gamma} with $\sqrt{\Omega_d}$ one has 
\begin{equation}
\label{eq:gamma-eta}
\gamma=\frac{\sqrt{\Omega_d}}{\sqrt{8\pi/3}}\cdot\eta_d
\end{equation}
with 
\begin{equation}
\eta_d=\frac{
\int\limits_{\varphi_1}^{\varphi_2} \int\limits_{\vartheta_1}^{\vartheta_2}\mathbf{p}_d\cdot\mathbf{E}_{f}
\sin\vartheta\ d\vartheta d\varphi
}
{
\sqrt{\Omega_d}
\cdot
\sqrt{\int\limits_{\varphi_1}^{\varphi_2} \int\limits_{\vartheta_1}^{\vartheta_2}|\mathbf{E}_f|^2
\sin\vartheta\ d\vartheta d\varphi}
}
\end{equation}
being the normalized overlap of the incident electric field with the desired dipole field
distribution computed solely on the parts of the focal sphere covered
by the incident light.
Insertion of Eq.~\ref{eq:gamma-eta} in Eq.~\ref{eq:field_gamma} yields
finally 
\begin{equation}
\label{eq:focusfield}
E_d(0)=-i\frac{\sqrt{2P}}{\lambda\sqrt{\epsilon_0c}} \cdot
\sqrt{\Omega_d}\cdot\eta_d \quad .
\end{equation}

This emphasizes that in order to maximize the focal field component
parallel to a given virtual dipole one has to (i) put as much of the
incident light as possible into the corresponding dipole mode and (ii)
maximize the weighted solid angle covered by the incident radiation.
It should be pointed out that the resulting field strength does not
depend on the dipole type.
Furthermore, Eq. \ref{eq:focusfield} constitutes the generalization of
Eqs. 8 and 10 of Ref. \cite{urbach2008}, where the case of a linear
dipole oriented along the optical axis of an objective obeying the
sine condition is treated.
However, the connection to dipole radiation was not established
there.

\section{Optimum irradiance distributions}
\label{sec:intensity}

Next, we want to derive the ideal radiation patterns incident onto the
focusing device which are transformed into a dipole wave that moves 
towards the focus.
For all devices treated here, we assume that they operate perfectly,
i.e., the wave incident onto the focus is assumed to have a uniform
phase after reflection/refraction off the focusing device (see also
Ref. \cite{bokor2008}).
For example, a parabolic mirror might exhibit (small) deviations from
the perfect parabolic shape.
Furthermore, the phase shift upon reflection is different for
different angles of incidence.
We presuppose that such effects are compensated for by, e.g.,
appropriate correcting elements. 
The production of such elements and the measurement of imperfections
might turn out to be a challenging but feasible experimental task
\cite{leuchs2008}. 

The method is simply to start with a dipole wave which virtually
emerges from the focus and trace it back through the optical element
under consideration.
We have chosen to formulate the ideal radiation patterns in terms of
irradiances instead of field vectors, since the former quantity is
the one usually measured in an experiment.
Furthermore, in the derivation of the apodization factors carried out
below, energy conservation considerations require the use of irradiances
and radiant intensities anyway.
The ideal vector fields can be constructed from the irradiance
patterns and the polarization patterns derived in the next section.

\begin{figure}
\centering
\includegraphics{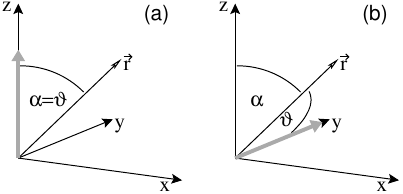}
\caption{\label{fig:coords}
Relation of the angles $\alpha$ and $\vartheta$ to the Cartesian
coordinates:
(a) dipole/quantization axis (gray arrow) parallel to the optical axis; 
(b) dipole/quantization axis perpendicular to the optical axis. 
}
\end{figure}

The derivation of the intensity patterns is pursued in the same way as
it was done in Ref. \cite{lindlein2007} for a parabolic mirror and a
linear dipole oriented parallel to the optical axis of the mirror.
We designate by $\alpha$ the angle that is enclosed by the optical
axis of the focusing element and the propagation direction of a single
ray which hits the optical element at a distance $r$ from the
optical axis.
Here, we chose the optical axis to coincide with the $z$-axis
(cf. Fig. \ref{fig:coords}).
Energy conservation demands that~\cite{lindlein2007}
\begin{equation}
\label{eq:apod}
\Upsilon(\alpha)\sin\alpha\ d\alpha= I(r)r\ dr\quad ,
\end{equation}
which delivers
\begin{equation}
\label{eq:irradiance}
I(r) = \Upsilon(\alpha)\cdot A(r,\alpha) \quad ,
\end{equation}
with $A(r,\alpha)=\frac{\sin\alpha(r)}{r}\
\frac{d\alpha(r)}{dr}$ being an apodization factor.
$\Upsilon(\alpha)$ is the radiant intensity, i.e., the light
power emitted into an infinitesimal solid angle.
$I(r)$ is the irradiance of the incident plane wave, i.e., the light
power incident onto an infinitesimal surface area.

In what follows, we express the apodization factor as a function of
the Cartesian coordinates $x,y$ in the plane perpendicular to the
optical axis: $A(x,y)=A(r(x,y))$ with
$r=\sqrt{x^2+y^2}$.
In the case of a parabolic mirror, $\alpha$ is given through the
relation \cite{lindlein2007}
\begin{equation}
\label{eq:angle_PM}
\tan\frac{\alpha}{2}=\frac{r}{2f} \quad,
\end{equation}
which with the help of some algebra leads to
\begin{equation}
A_{PM}(x,y)=\frac{1}{f^2\left(\frac{x^2}{4f^2}+\frac{y^2}{4f^2}+1\right)^2}
\quad .
\end{equation} 
For an aplanatic lens that fulfills the sine condition, i.e., a lens
where the refracted rays of same phase seem to emerge from a spherical
surface around the focus, one has 
\begin{equation}
\label{eq:angle_SC}
\sin\alpha=\frac{r}{f} \quad,
\end{equation}
resulting in 
\begin{equation}
A_{AL}(x,y)=\frac{1}{f^2\sqrt{1-\frac{x^2}{f^2}-\frac{y^2}{f^2}}}
\quad .
\end{equation} 
As a third focusing element, we consider an ideal thin lens, i.e., a
lens for which the refracted rays seem to emerge from a plane
perpendicular to the optical axis.
A practical realization of such a thin lens would be an idealized
diffractive optical lens (e.g., as used recently for imaging ion
fluorescence~\cite{streed2011}). 
One has
\begin{equation}
\label{eq:angle_TE}
\tan\alpha=\frac{r}{f} \quad,
\end{equation}
which delivers
\begin{equation}
A_{TL}(x,y)=\frac{1}{f^2\left(\frac{x^2}{f^2}+\frac{y^2}{f^2}+1\right)^{3/2}}
\quad .
\end{equation} 

Next, for every point $(x,y)$ in the plane perpendicular to the optical
axis we determine $\vartheta$.
The radiant intensity along this direction is given by the
angular dipole radiation pattern $D(\vartheta)$:
\begin{equation}
\Upsilon(\vartheta)=D_0\cdot D(\vartheta)
\quad,
\end{equation}
where $D_0$ is a proportionality constant.
Finally, $\Upsilon(\vartheta)$ is also expressed in terms of $x,y$ and
multiplied with the apodization factor $A(x,y)$ yielding the irradiance
$I(x,y)$.


\begin{figure}[t]
\centering
\includegraphics{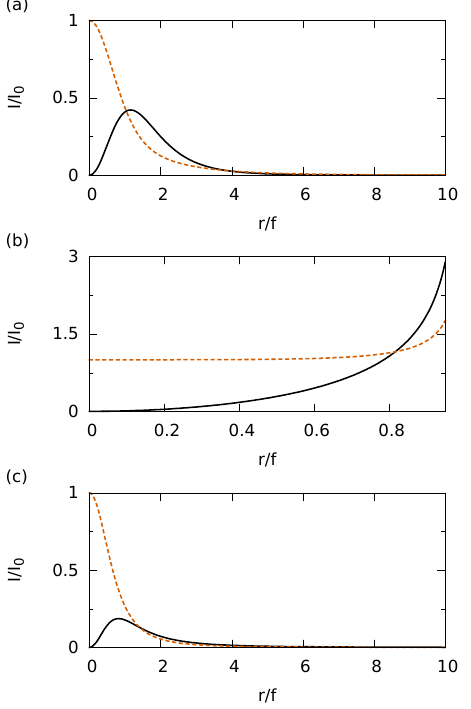}
\caption{\label{fig:para} 
Ideal irradiance distributions for dipoles oriented parallel to the
optical axis of the focusing device: (a) parabolic mirror, (b)
aplanatic lens fulfilling the sine condition, (c) ideal thin lens.
Solid (dashed) lines denote the irradiance for a linear (circular)
dipole radiation pattern.
}
\end{figure}

If the dipole is oriented parallel to the optical axis, $\alpha$
coincides with $\vartheta$ (see Fig. \ref{fig:coords}a).
Hence, the ideal irradiation is rotationally symmetric with
respect to the optical axis and the angular dipole emission patterns
given above can be written as
$D_{\pi,\sigma_\pm}(\vartheta)=D_{\pi,\sigma_\pm}(\alpha)$.
Using Eqs. \ref{eq:angle_PM}, \ref{eq:angle_SC} and \ref{eq:angle_TE},
respectively, and applying several trigonometric relations one obtains
the ideal irradiance distributions.
They are summarized in Tab. \ref{tab:para} and plotted in
Fig. \ref{fig:para}.
We have also introduced normalized coordinates $X=x/f$, $Y=y/f$,
$R=r/f$.
All remaining dimensional quantities have been lumped into a
proportionality constant $I_0$.

\begin{table*}
\caption{\label{tab:para}
  Optimum irradiance distributions $I(R)$ for the creation of
  radiation patterns of dipoles oriented parallel to the optical axis
  of the focusing device. $R=r/f$ is the radial coordinate transverse to the
  optical axis given in units of the focal length. The proportionality
  constant $I_0$ contains all dimensional quantities.}

\begin{tabular}{c|c|c|c}
& parabolic mirror & aplanatic lens & ideal thin lens \\
\hline
linear dipole & 
\begin{minipage}[b]{.2\textwidth}
$$I_0\cdot\frac{R^2}{(\frac{R^2}{4}+1)^4}$$
\end{minipage}
& 
\begin{minipage}[b]{.2\textwidth}
 $$I_0\cdot\frac{R^2}{\sqrt{1-R^2}}$$
\end{minipage}
& 
\begin{minipage}[b]{.2\textwidth}
$$I_0\cdot\frac{R^2}{(R^2+1)^{\frac{5}{2}}}$$
\end{minipage}
\\[2em]
\hline
circular dipole & 
\begin{minipage}[b]{.2\textwidth}
$$I_0\cdot\frac{\frac{R^4}{16}+1}{(\frac{R^2}{4}+1)^4}$$
\end{minipage}
& 
\begin{minipage}[b]{.2\textwidth}
$$I_0\cdot\frac{1-\frac{R^2}{2}}{\sqrt{1-R^2}}$$
\end{minipage}
&
\begin{minipage}[b]{.2\textwidth}
$$I_0\cdot\frac{1+\frac{R^2}{2}}{(R^2+1)^{\frac{5}{2}}}$$
\end{minipage}
\\
\end{tabular}
\end{table*}

The formula obtained here for the irradiance in the case of an
aplanatic lens fulfilling the sine condition and a linear dipole
corresponds exactly to the square of the formula for the electric
field amplitude in the entrance pupil of the lens as it was found by
means of an optimization method in Ref. \cite{urbach2008}.


\begin{figure}
\centering
\includegraphics{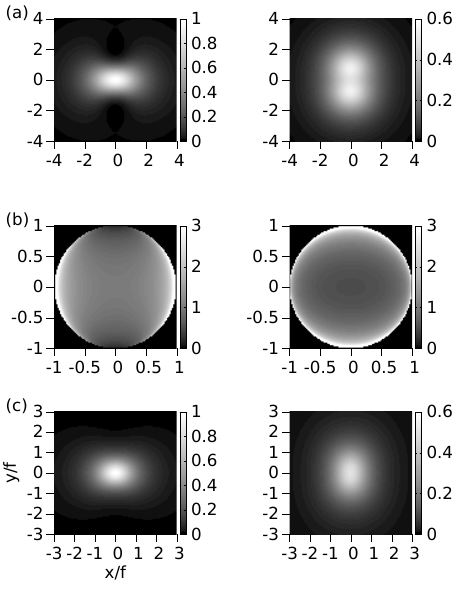}
\caption{\label{fig:ortho} 
Ideal irradiance distributions for dipoles oriented orthogonal to the
optical axis of the focusing device: (a) parabolic mirror, (b) lens
fulfilling the sine condition, (c) ideal thin lens.
Left panel: linear dipole; right panel: circular dipole.
The gray levels designate the values of $I/I_0$.
}
\end{figure}

Next, we turn to the case of the virtual dipole being oriented
perpendicular to the optical axis of the focusing element.
Without loss of generality we choose the virtual dipole
axis (the direction $\vartheta$=0) to coincide with the $y$-axis (see
Fig. \ref{fig:coords}b). 
An immediate consequence of the dipole axis being orthogonal to the
optical axis is that the ideal intensity distributions have no longer
rotational symmetry with respect to the optical axis.
Furthermore, the angles $\alpha$ and $\vartheta$ do not coincide any
more.
Thus, the relations given by Eqs. \ref{eq:angle_PM}, \ref{eq:angle_SC}
and \ref{eq:angle_TE} cannot be used to express the angular emission
patterns $D_{\pi,\sigma_\pm}(\vartheta)$ in terms of the transverse
coordinates.
Instead one has to use in the case of the parabolic mirror the
relations  
\begin{equation}
\cos\vartheta=\frac{y}{\sqrt{x^2+y^2+z^2}}
\quad ,
z=f-\frac{x^2+y^2}{4f}
\quad. 
\end{equation}
This is valid if the focus of the parabolic mirror is located in the
origin of the coordinate system.
For a lens obeying the sine condition the relation between
$\vartheta$ and $x,y$ reads 
\begin{equation}
\cos\vartheta=\frac{y}{f} \quad ,
\end{equation}
and for the ideal thin lens one has
\begin{equation}
\tan\vartheta=\frac{\sqrt{x^2+f^2}}{y} \quad .
\end{equation}
Again, the focus is located in the origin of the coordinate system.
Using some trigonometric relations and multiplication with the
corresponding apodization factors leads to the ideal irradiance
distributions given in Tab. \ref{tab:ortho} and displayed in
Fig. \ref{fig:ortho}. 

\begin{table*}
\caption{\label{tab:ortho}
  Optimum irradiance distributions $I(X,Y)$ for the creation of
  radiation patterns of dipoles oriented orthogonal to the optical axis
  of the focusing device. $X=x/f$ and $Y=y/f$ are the coordinates
  transverse to the optical axis given in units of the focal
  length. The proportionality constant $I_0$ contains all dimensional
  quantities.}

\begin{tabular}{c|c|c|c}
& parabolic mirror & aplanatic lens & ideal thins lens \\
\hline
linear dipole & 
\begin{minipage}[b]{.35\textwidth}
$$I_0\cdot\frac{1-\frac{Y^2}{X^2+Y^2+(X^2/4+Y^2/4-1)^2}}
{(X^2/4+Y^2/4+1)^2}$$
\end{minipage}
& 
\begin{minipage}[b]{.22\textwidth}
$$I_0\cdot\frac{1-Y^2}{\sqrt{1-X^2-Y^2}}$$
\end{minipage}
& 
\begin{minipage}[b]{.22\textwidth}
$$I_0\cdot\frac{1+X^2}{(X^2+Y^2+1)^{5/2}}$$
\end{minipage}
\\[2em]
\hline
circular dipole & 
\begin{minipage}[b]{.3\textwidth}
$$I_0\cdot\frac{1/2+\frac{Y^2/2}{X^2+Y^2+(X^2/4+Y^2/4-1)^2}}
{(X^2/4+Y^2/4+1)^2}$$
\end{minipage}
& 
\begin{minipage}[b]{.2\textwidth}
$$I_0\cdot\frac{1/2+Y^2/2}{\sqrt{1-X^2-Y^2}}$$
\end{minipage}
&
\begin{minipage}[b]{.2\textwidth}
$$I_0\cdot\frac{Y^2+\frac{X^2+1}{2}}{(X^2+Y^2+1)^{5/2}}$$
\end{minipage}
\\
\end{tabular}
\end{table*}

\section{Optimum states of polarization}
\label{sec:polarization}

We begin the discussion for a virtual dipole parallel to the
optical axis.
For a linear dipole, the polarization pattern can be obtained
immediately from the fact that the polarization vector of a linear
dipole is oriented along $\mathbf{e}_\vartheta$ in the far field.
This results in the well known radial polarization
pattern \cite{quabis2000,lindlein2007,bokor2004} and, of course,
coincides with the result found recently by optimization
\cite{urbach2008}.
The polarization pattern is the same for all of the devices discussed
here.

The situation is slightly more complex in the case of a circular
dipole.
The polarization vector for a certain $\vartheta$ is given by
$\mathbf{p}_{\sigma_\pm}$ in Eq. \ref{eq:polvec_circ}.
Using the relations given in Ref.~\cite{mandel-wolf1995} the
normalized Stokes parameter $S_3/S_0$ describing the degree of
circular polarization is readily determined to be
\begin{equation}
\frac{{S_3}_{\sigma_\pm}}{{S_0}_{\sigma_\pm}}=
\frac{\pm 2\cos\vartheta}{1+\cos^2\vartheta}\quad,
\end{equation} 
where $S_0$ corresponds to the total intensity and $S_3$ to the
intensity that would be measured after passing the beam through a
circular polarizer.
With $S_3/S_0=\sin(2\chi)$ giving the so called ellipticity angle
\cite{hecht1987}, the ellipticity as a function of $\vartheta$ is
\begin{equation}
\chi_{\sigma_\pm}=\frac{\arcsin\left(
\frac{\pm 2\cos\vartheta}{1+\cos^2\vartheta}\right)
}{2}\quad ,
\end{equation}
where $\chi_{\sigma_\pm}=\pm\pi/4$ corresponds to circular states of
polarization. 
This shows that the ideal polarization pattern for circular dipoles
exhibits a varying ellipticity as a function of the radial distance to
the optical axis.

\begin{figure}
\centering
\includegraphics{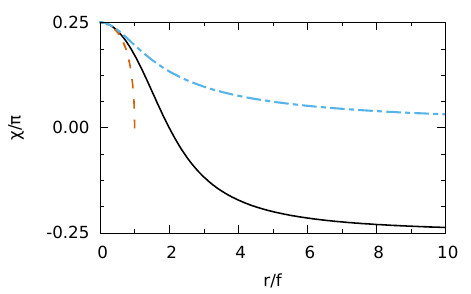}
\caption{\label{fig:ellip_para}
Ellipticity of the ideal states of polarization for a circular dipole
($\sigma_+$, a sign change gives the values for
$\sigma_-$) oriented parallel to the optical axis of the focusing
device.
Solid line: parabolic mirror; dashed line: aplanatic lens; dash-dotted 
line: ideal thin lens.
}
\end{figure}

Using Eqs. \ref{eq:angle_PM}, \ref{eq:angle_SC} and \ref{eq:angle_TE}
to express $\vartheta$ through $r$, one obtains the ideal state of
polarization for the different focusing devices.
The results are displayed in Fig. \ref{fig:ellip_para}.
Note that in the case of the parabolic mirror there is a sign change
of the ellipticity at the radius $r=2f$, which corresponds to a half
opening angle of the parabolic mirror of 90$^\circ$. 
This result explains the finding of  Ref. \cite{bokor2008}(Fig. 2)
that an increase of the half focusing angle beyond 90$^\circ$ does not
lead to a substantial decrease of focus size for an incident wave that
has a circular state of polarization of same helicity over the whole
beam cross section. 

\begin{figure}
\centering
\includegraphics{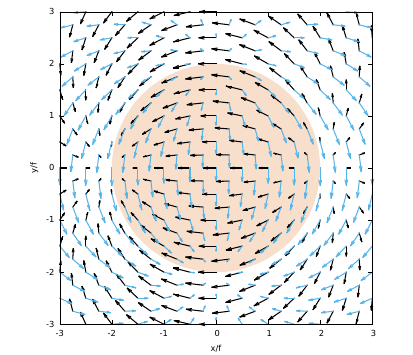}
\caption{\label{fig:pol_PM_para}
State of polarization of the ideal radiation pattern for the case of
a parabolic mirror and a circular dipole ($\sigma_+$) with the dipole
axis oriented parallel to the optical axis of the mirror.
The vectors in light blue color denote the linearly polarized
components with $\pi/2$ phase shift.
The shaded area denotes the region of NA=1.
}
\end{figure}

The spatial polarization pattern in the entrance plane of the
focusing devices is calculated exemplarily for a parabolic
mirror.
First, $\mathbf{p}_{\sigma_\pm}$ is transformed into a unit vector
$\hat{\mathbf{p}}_{\sigma_\pm}$ by multiplication with
$\sqrt{2}/\sqrt{1+\cos^2\vartheta}$. 
Then, the polarization pattern after reflection off the parabolic
mirror is obtained as follows:
The state of polarization $\hat{\mathbf{p}}_{\vartheta,\varphi}$ of the
virtual dipole, formulated in terms of the spherical unit vectors
$\mathbf{e}_{\vartheta,\varphi}$, is transformed into the Cartesian
coordinate system:
$\hat{\mathbf{p}}_{\vartheta,\varphi}\rightarrow\hat{\mathbf{p}}_{x,y,z}$.
Then, the ray for each emission direction is traced from the virtual
dipole towards the surface of the parabolic mirror.
The polarization vector of the ray after reflection at the mirror
surface is calculated via the following equation:
\begin{equation}
  \mathbf{\hat{p}}= 2
  (\hat{\mathbf{p}}_{x,y,z}\cdot
  \mathbf{N}) \mathbf{N} -  \hat{\mathbf{p}}_{x,y,z}
\quad ,
\end{equation}
with the surface normal of the mirror given by
\begin{equation}
\mathbf{N}=\frac{1}{\sqrt{\frac{x^2+y^2}{4f^2}+1}}
\left(
\begin{array}{c}
\frac{x}{2f}\\
\frac{y}{2f}\\
1
\end{array}
\right)
\quad .
\end{equation}
Again, the focus is chosen to be at the origin of the
coordinate system.
For $\sigma_\pm$ light, this procedure results in
\begin{equation}
\begin{split}
\hat{\mathbf{p}}_{\sigma_\pm}(X,Y)=&\frac{1}
{\sqrt{2\left[(X^2+Y^2)^2+16\right]}}\times\\
&
\left[
\left(
\begin{array}{c}
X^2-Y^2-4\\
2XY\end{array}
\right)
\pm i
\left(
\begin{array}{c}
2XY\\
Y^2-X^2-4
\end{array}
\right)
\right]
\end{split}
\quad .
\end{equation}
This pattern is depicted in Fig. \ref{fig:pol_PM_para}. 
Up to the boundary given by $\sqrt{X^2+Y^2}=2$ (or numerical aperture
NA=1), the polarization pattern is qualitatively similar for
the other focusing devices treated here. 
However, one has to keep in mind that the relations between
$\vartheta$, $\varphi$ and the Cartesian coordinates in the entrance
plane of the devices are different.
Thus, the pattern displayed inside the circle marking NA=1 in
Fig.~\ref{fig:pol_PM_para} will be distorted or stretched/compressed
to some amount, respectively.

\begin{figure}
\centering
\includegraphics{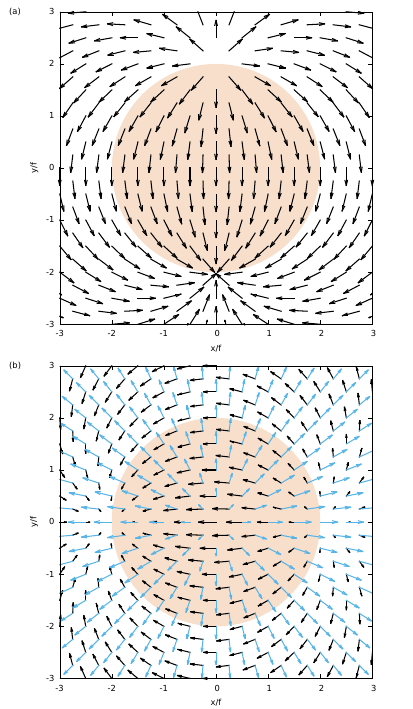}
\caption{\label{fig:pol_PM_ortho}
State of polarization of the ideal radiation patterns for the case of
a parabolic mirror and the dipole axis oriented orthogonal to the
optical axis of the mirror: (a) linear dipole, (b) circular dipole
($\sigma_+$).
In (b) the vectors in light blue color denote the linearly polarized
components with $\pi/2$ phase shift.
The shaded area denotes the region of NA=1.
}
\end{figure}

If the dipole axis is orthogonal to the optical axis, the procedure is
the same except that in this case the direction $\vartheta=0$
coincides with the positive $y$-axis. 
This is handled by interchanging the second with the third component
of the unit vectors $\mathbf{e}_{\vartheta,\varphi}$ of the spherical
coordinate system.
Using Eqs. \ref{eq:polvec_lin} and \ref{eq:polvec_circ}, one obtains
the polarization patterns
\begin{equation}
\hat{\mathbf{p}}_\pi(X,Y)=\frac{1}{\sqrt{(X^2+Y^2+4)^2-16Y^2}}
\left(
\begin{array}{c}
2XY\\
Y^2-X^2-4
\end{array}
\right)
\quad 
\end{equation}
for a linear dipole
and
\begin{equation}
\begin{split}
\hat{\mathbf{p}}_{\sigma_\pm}(X,Y)=&
\frac{1}{\sqrt{16Y^2+(4+X^2+Y^2)^2}}\times\\
&
\left[
\left(
\begin{array}{c}
X^2-Y^2-4\\
2XY
\end{array}
\right)
 \pm i
 \left(
 \begin{array}{c}
4X\\
4Y
\end{array}
 \right)
\right]
\end{split}
\end{equation}
for the circular dipoles.
The two polarization patterns are illustrated in
Fig.~\ref{fig:pol_PM_ortho}.

\section{Overlap of practical distributions with the ideal ones}
\label{sec:overlap}

Some of the radiation patterns derived above -- especially in the case of
the dipole axis parallel to the optical axis -- exhibit obvious
similarities with radiation modes which are commonly used in
experiments.  
In the case of a parabolic mirror and the ideal thin lens,
the irradiance distributions for a linear virtual dipole look close to
the well known doughnut modes.
In the case of circular virtual dipoles there is a strong similarity
with fundamental Gaussian distributions.

In order to estimate and quantify the strength of the similarities, we
will calculate the overlap of these modes with the ideal radiation
distributions.
We begin with recalling that the transformation of the field
distribution in the entrance plane of the focusing device to the one
on the focal sphere is of course the same for the ideal dipole fields
and the incident fields.
Therefore, it does not matter whether the overlap $\eta_d$ in
Eq.~\ref{eq:focusfield} is calculated on the focal sphere or in the
entrance aperture of the focusing optics. 
Since the field in the entrance aperture is the one that is more easily
accessible in experiments, all overlaps are calculated there in what
follows.

\begin{figure}
\centering
\includegraphics{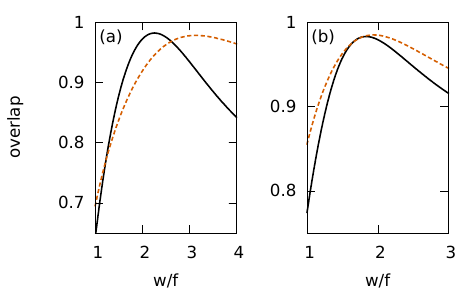}
\caption{\label{fig:ovl_examples}
Normalized overlap of the ideal field distributions with common
radiation modes as a function of the modes beam radius $w$ in units of
the focal length $f$.
The virtual dipoles are oriented parallel to the optical axis.
(a) Parabolic mirror with a half opening angle of 134$^\circ$.
(b) Ideal thin lens with a numerical aperture of 0.95.
Solid lines: Overlap of the ideal field for linear dipoles with a
radially polarized doughnut mode.
Dashed lines: Overlap of the ideal field for circular dipoles with a
fundamental Gaussian mode that has an ideal state of polarization.
}
\end{figure}

Figure \ref{fig:ovl_examples} shows some examples for illustration.
In the case of a parabolic mirror with a half opening angle of 134$^\circ$
as it is used in the experiment described in
Refs. \cite{lindlein2007,sondermann2007}, a radially polarized
doughnut mode ($\sqrt{I(r)}\sim r\cdot\textnormal{e}^{-r^2/w^2}$, $w$
is the beam radius) best matches the field distribution for a linear
dipole parallel to the optical axis at a   beam radius of $w/f=$2.26.
There, the overlap is $\eta_{\pi}=$0.982.
The half opening angle of 134$^\circ$ corresponds to
$\Omega_\pi=$0.94$\times8\pi/3$.
Thus, the field amplitude parallel to the virtual linear dipole would
be 0.95 times the one obtained with the ideal radiation pattern and a
parabolic mirror of infinite dimensions.
This has been confirmed in simulations. 
In the case of a circular dipole parallel to the optical axis, the
maximum overlap of $\eta_{\sigma_\pm}=$0.978 is achieved for a
Gaussian fundamental mode ($\sqrt{I(r)}\sim\textnormal{e}^{-r^2/w^2}$)
with a beam waist of $w/f=$3.14. 
With $\Omega_{\sigma_\pm}=$0.80$\times8\pi/3$ one achieves a field
amplitude parallel to the virtual circular dipole of 0.87 times the
maximum possible one.

For an ideal thin lens and dipoles parallel to the optical axis
(Fig. \ref{fig:ovl_examples}b) with NA=0.95 the 
optimum overlap of $\eta_{\pi}=$0.983 is reached at $w/f=$1.84
(linear dipole) and the maximum value of $\eta_{\sigma_\pm}=$0.985
reached at $w/f=$1.94 (circular dipole).
With $\Omega_\pi=$0.27$\times8\pi/3$ and
$\Omega_{\sigma_\pm}=$0.38$\times8\pi/3$ one achieves 0.51 and
0.61  times the field amplitudes of the maximum possible ones,
respectively. 

Please note that for the parabolic mirror as well as for the ideal thin lens
the optimum beam waist of the doughnut is such that the maximum of the
irradiance distribution is located at a somewhat larger radial position than
the maximum of the radiation patterns of a linear dipole that are
displayed in Fig. \ref{fig:para}(a,c).
Furthermore, the values for the circular dipoles given above have to
be considered as upper limits in the sense that the ideal state of
polarization for a circular dipole is difficult to obtain in practice
(see below).

\begin{figure}
\centering
\includegraphics{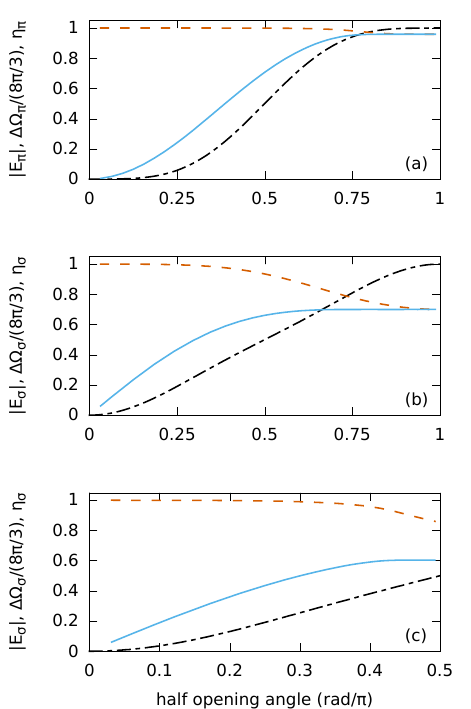}
\caption{\label{fig:ovl_sa}
Weighted solid angle $\Omega_d/(8\pi/3)$ (dash-dotted line), maximum
possible overlap $\eta_{d}$ with a linear dipole wave (dashed
line) and field amplitude $|E_d|$ in the focus in units of
the maximum possible one (solid line) as a function of the half
opening angle for (a) a parabolic mirror illuminated by a radially
polarized doughnut mode, (b) a parabolic mirror illuminated by a
circularly polarized fundamental Gaussian and (c) an ideal thin lens
illuminated by a circularly polarized fundamental Gaussian.
For further explanations see text.
}
\end{figure}

If the weighted solid angle covered by the focusing optics is increased, the
maximum achievable overlap of the 'simple' modes with the ideal
distributions decreases.
However, this is overcompensated for by the increase in weighted solid
angle, i.e., the field in the focus parallel to a specific virtual
dipole is maximized by maximizing the weighted solid angle.
This is illustrated in Fig. \ref{fig:ovl_sa}(a) for a linear
dipole parallel to the optical axis and a parabolic mirror
(illuminated by a radially polarized doughnut).
For each (half) opening angle of the mirror the maximum overlap is
obtained from a curve like the one in Fig. \ref{fig:ovl_examples}(a).
The resulting field is then obtained from Eq. \ref{eq:focusfield}
by inserting the overlap value and the corresponding square root of
the covered weighted solid angle into this equation.

From the asymptotic behavior of the plot for the field amplitude in
Fig. \ref{fig:ovl_sa}(a) one can extract the maximum field
amplitude one could achieve with a doughnut mode of suitable
polarization by focusing the mode with a parabolic mirror.
Likewise, one obtains the same values for circular dipoles
(i.e., overlapping with a Gaussian mode of proper polarization) and/or
an ideal thin lens.
The values are listed in Tab. \ref{tab:max_fields}.

\begin{table}
\caption{\label{tab:max_fields}
Maximum electric field amplitudes 
achievable by use of a radially polarized doughnut mode and a
fundamental Gaussian mode with the state of polarization matching a
circular dipole.
The values are given relative to the ones obtainable with the ideal
irradiance distributions of Fig. \ref{fig:para}(a,c).
The virtual dipoles are oriented parallel to the optical axis.
}
\begin{tabular}{c|c|c}
& doughnut mode & fundamental Gaussian\\
\hline 
parabolic mirror &  0.958 &  0.919 \\
\hline
ideal thin lens & 0.575 & 0.628\\
\end{tabular}
\end{table}

So far, we have assumed modes with an ideal state of
polarization. 
As an illustration we now treat the fields obtainable with circularly
polarized fundamental Gaussian modes.
For a parabolic mirror and a circular dipole parallel to the optical
axis the maximum achievable overlap drops significantly at about half
opening angles of $\pi/2$ (cf. Fig. \ref{fig:ovl_sa}b).
Likewise, the corresponding field in the focal point saturates at this
value.
We attribute this behavior to the change of the helicity of the ideal
polarization pattern at $\pi/2$ half opening angle.
In other words, the uniformly circular polarized beam is strongly
mismatched to the ideal pattern at larger angles.
For full solid angle, the field strength approaches 0.7 times the
maximum possible one.
Hence, the imperfect polarization state drastically limits the
attainable field strength and an increase of the mirror size beyond
half solid angle does not result in a significant gain of the focal
field amplitude.

Figure \ref{fig:ovl_sa}(c) treats the case of an ideal thins lens
that is illuminated by a circularly polarized fundamental Gaussian
beam.
If the NA approaches a value of $1$, the maximum
achievable field tends to 0.603 times the one for a perfect dipole
wave. 
This value corresponds perfectly to the one found in
Ref.~\cite{tey2009} for the same setup. 
Moreover, the corresponding beam radius that gives the best overlap is
exactly the same that is found there.

For the case of a lens obeying the sine condition  the calculation of
the overlap with the simple modes discussed above does not appear to
be meaningful on first sight.
This is due to the divergence of the ideal irradiance values towards
the boundary of the lens aperture.
Nevertheless, large overlaps can be achieved if the beam radius tends
towards values larger than the aperture size.
For example, for an objective with NA=0.95 the overlap of a radially
polarized doughnut mode with the ideal dipole field is 0.927 if the beam
radius matches the aperture radius.
However, already for this value roughly 40\% of the power contained in
the doughnut mode is cut off by the aperture boundary.
At even larger beam radii the overlap saturates at a value of about
0.989 by the cost of losing practically all photons contained in the
doughnut mode at the lens aperture.

Likewise, the overlap of a linearly polarized fundamental Gaussian
mode with the ideal field of a linear dipole oriented
\emph{orthogonal} to the optical axis increases continuously with beam
radius for an aplanatic lens.
For NA=0.95 the overlap is 0.927 if beam radius and aperture radius
are equal (14\% loss of power by cutting off the tail of the Gaussian
mode).
The overlap for much larger beam radii is saturating at about 0.972. 

\section{Concluding remarks}
\label{sec:remarks}

We have derived the ideal radiation patterns with which one achieves
the maximum possible field strengths for several focusing devices.
Besides using these radiation patterns, it is of utmost importance to
maximize the covered weighted solid angle, as it is shown by
Eq.~\ref{eq:focusfield}.
Thus, the maximum possible electric field strength is achieved by
covering the full solid angle.
Also, the maximum possible effect in the interaction of light with
single atoms will also only be found for full solid angle coverage,
since scattering as well as absorption of photons is proportional to
the square of the electric field amplitude at the location of the
atom~\cite{mandel-wolf1995}.
According with the papers of Van Enk and Kimble
\cite{vanenk2004,vanenk2001}, we want to stress once more that for
optimum interaction of light with single atoms one has to shape the
radiation incident onto the atom to resemble an electric dipole wave.



\begin{thebibliography}{34}
\expandafter\ifx\csname natexlab\endcsname\relax\def\natexlab#1{#1}\fi
\expandafter\ifx\csname bibnamefont\endcsname\relax
  \def\bibnamefont#1{#1}\fi
\expandafter\ifx\csname bibfnamefont\endcsname\relax
  \def\bibfnamefont#1{#1}\fi
\expandafter\ifx\csname citenamefont\endcsname\relax
  \def\citenamefont#1{#1}\fi
\expandafter\ifx\csname url\endcsname\relax
  \def\url#1{\texttt{#1}}\fi
\expandafter\ifx\csname urlprefix\endcsname\relax\def\urlprefix{URL }\fi
\providecommand{\bibinfo}[2]{#2}
\providecommand{\eprint}[2][]{\url{#2}}

\bibitem[{\citenamefont{Vamivakas et~al.}(2007)\citenamefont{Vamivakas,
  Atat\"ure, Dreiser, Yilmaz, Badolato, Swan, Goldberg, Imamoglu, and
  \"Unl\"u}}]{vamivakas2007}
\bibinfo{author}{\bibfnamefont{A.~N.} \bibnamefont{Vamivakas}},
  \bibinfo{author}{\bibfnamefont{M.}~\bibnamefont{Atat\"ure}},
  \bibinfo{author}{\bibfnamefont{J.}~\bibnamefont{Dreiser}},
  \bibinfo{author}{\bibfnamefont{S.~T.} \bibnamefont{Yilmaz}},
  \bibinfo{author}{\bibfnamefont{A.}~\bibnamefont{Badolato}},
  \bibinfo{author}{\bibfnamefont{A.~K.} \bibnamefont{Swan}},
  \bibinfo{author}{\bibfnamefont{B.~B.} \bibnamefont{Goldberg}},
  \bibinfo{author}{\bibfnamefont{A.}~\bibnamefont{Imamoglu}}, \bibnamefont{and}
  \bibinfo{author}{\bibfnamefont{M.~S.} \bibnamefont{\"Unl\"u}},
  \bibinfo{journal}{Nano Letters} \textbf{\bibinfo{volume}{7}},
  \bibinfo{pages}{2892} (\bibinfo{year}{2007}).

\bibitem[{\citenamefont{Wrigge et~al.}(2008)\citenamefont{Wrigge, Gerhardt,
  Hwang, Zumofen, and Sandoghdar}}]{wrigge2008}
\bibinfo{author}{\bibfnamefont{G.}~\bibnamefont{Wrigge}},
  \bibinfo{author}{\bibfnamefont{I.}~\bibnamefont{Gerhardt}},
  \bibinfo{author}{\bibfnamefont{J.}~\bibnamefont{Hwang}},
  \bibinfo{author}{\bibfnamefont{G.}~\bibnamefont{Zumofen}}, \bibnamefont{and}
  \bibinfo{author}{\bibfnamefont{V.}~\bibnamefont{Sandoghdar}},
  \bibinfo{journal}{Nature Physics} \textbf{\bibinfo{volume}{4}},
  \bibinfo{pages}{60} (\bibinfo{year}{2008}).

\bibitem[{\citenamefont{Tey et~al.}(2008)\citenamefont{Tey, Chen, Aljunid,
  Chng, Huber, Maslennikov, and Kurtsiefer}}]{tey2008-np}
\bibinfo{author}{\bibfnamefont{M.~K.} \bibnamefont{Tey}},
  \bibinfo{author}{\bibfnamefont{Z.}~\bibnamefont{Chen}},
  \bibinfo{author}{\bibfnamefont{S.~A.} \bibnamefont{Aljunid}},
  \bibinfo{author}{\bibfnamefont{B.}~\bibnamefont{Chng}},
  \bibinfo{author}{\bibfnamefont{F.}~\bibnamefont{Huber}},
  \bibinfo{author}{\bibfnamefont{G.}~\bibnamefont{Maslennikov}},
  \bibnamefont{and}
  \bibinfo{author}{\bibfnamefont{C.}~\bibnamefont{Kurtsiefer}},
  \bibinfo{journal}{Nature Physics} \textbf{\bibinfo{volume}{4}},
  \bibinfo{pages}{924} (\bibinfo{year}{2008}).

\bibitem[{\citenamefont{Zumofen et~al.}(2008)\citenamefont{Zumofen, Mojarad,
  Sandoghdar, and Agio}}]{zumofen2008}
\bibinfo{author}{\bibfnamefont{G.}~\bibnamefont{Zumofen}},
  \bibinfo{author}{\bibfnamefont{N.~M.} \bibnamefont{Mojarad}},
  \bibinfo{author}{\bibfnamefont{V.}~\bibnamefont{Sandoghdar}},
  \bibnamefont{and} \bibinfo{author}{\bibfnamefont{M.}~\bibnamefont{Agio}},
  \bibinfo{journal}{Phys. Rev. Lett.} \textbf{\bibinfo{volume}{101}},
  \bibinfo{pages}{180404} (\bibinfo{year}{2008}).

\bibitem[{\citenamefont{Tey et~al.}(2009)\citenamefont{Tey, Maslennikov, Liew,
  Aljunid, Huber, Chng, Chen, Scarani, and Kurtsiefer}}]{tey2009}
\bibinfo{author}{\bibfnamefont{M.~K.} \bibnamefont{Tey}},
  \bibinfo{author}{\bibfnamefont{G.}~\bibnamefont{Maslennikov}},
  \bibinfo{author}{\bibfnamefont{T.~C.~H.} \bibnamefont{Liew}},
  \bibinfo{author}{\bibfnamefont{S.~A.} \bibnamefont{Aljunid}},
  \bibinfo{author}{\bibfnamefont{F.}~\bibnamefont{Huber}},
  \bibinfo{author}{\bibfnamefont{B.}~\bibnamefont{Chng}},
  \bibinfo{author}{\bibfnamefont{Z.}~\bibnamefont{Chen}},
  \bibinfo{author}{\bibfnamefont{V.}~\bibnamefont{Scarani}}, \bibnamefont{and}
  \bibinfo{author}{\bibfnamefont{C.}~\bibnamefont{Kurtsiefer}},
  \bibinfo{journal}{New Journal of Physics} \textbf{\bibinfo{volume}{11}},
  \bibinfo{pages}{043011} (\bibinfo{year}{2009}).

\bibitem[{\citenamefont{Slodi\ifmmode~\check{c}\else \v{c}\fi{}ka
  et~al.}(2010)\citenamefont{Slodi\ifmmode~\check{c}\else \v{c}\fi{}ka,
  H\'etet, Gerber, Hennrich, and Blatt}}]{slodicka2010}
\bibinfo{author}{\bibfnamefont{L.}~\bibnamefont{Slodi\ifmmode~\check{c}\else
  \v{c}\fi{}ka}}, \bibinfo{author}{\bibfnamefont{G.}~\bibnamefont{H\'etet}},
  \bibinfo{author}{\bibfnamefont{S.}~\bibnamefont{Gerber}},
  \bibinfo{author}{\bibfnamefont{M.}~\bibnamefont{Hennrich}}, \bibnamefont{and}
  \bibinfo{author}{\bibfnamefont{R.}~\bibnamefont{Blatt}},
  \bibinfo{journal}{Phys. Rev. Lett.} \textbf{\bibinfo{volume}{105}},
  \bibinfo{pages}{153604} (\bibinfo{year}{2010}).

\bibitem[{\citenamefont{Lindlein et~al.}(2007)\citenamefont{Lindlein, Maiwald,
  Konermann, Sondermann, Peschel, and Leuchs}}]{lindlein2007}
\bibinfo{author}{\bibfnamefont{N.}~\bibnamefont{Lindlein}},
  \bibinfo{author}{\bibfnamefont{R.}~\bibnamefont{Maiwald}},
  \bibinfo{author}{\bibfnamefont{H.}~\bibnamefont{Konermann}},
  \bibinfo{author}{\bibfnamefont{M.}~\bibnamefont{Sondermann}},
  \bibinfo{author}{\bibfnamefont{U.}~\bibnamefont{Peschel}}, \bibnamefont{and}
  \bibinfo{author}{\bibfnamefont{G.}~\bibnamefont{Leuchs}},
  \bibinfo{journal}{Laser Physics} \textbf{\bibinfo{volume}{17}},
  \bibinfo{pages}{927} (\bibinfo{year}{2007}).

\bibitem[{\citenamefont{Sondermann et~al.}(2007)\citenamefont{Sondermann,
  Maiwald, Konermann, Lindlein, Peschel, and Leuchs}}]{sondermann2007}
\bibinfo{author}{\bibfnamefont{M.}~\bibnamefont{Sondermann}},
  \bibinfo{author}{\bibfnamefont{R.}~\bibnamefont{Maiwald}},
  \bibinfo{author}{\bibfnamefont{H.}~\bibnamefont{Konermann}},
  \bibinfo{author}{\bibfnamefont{N.}~\bibnamefont{Lindlein}},
  \bibinfo{author}{\bibfnamefont{U.}~\bibnamefont{Peschel}}, \bibnamefont{and}
  \bibinfo{author}{\bibfnamefont{G.}~\bibnamefont{Leuchs}},
  \bibinfo{journal}{Appl. Phys. B} \textbf{\bibinfo{volume}{89}},
  \bibinfo{pages}{489} (\bibinfo{year}{2007}).

\bibitem[{\citenamefont{Pinotsi and Imamoglu}(2008)}]{pinotsi2008}
\bibinfo{author}{\bibfnamefont{D.}~\bibnamefont{Pinotsi}} \bibnamefont{and}
  \bibinfo{author}{\bibfnamefont{A.}~\bibnamefont{Imamoglu}},
  \bibinfo{journal}{Phys. Rev. Lett.} \textbf{\bibinfo{volume}{100}},
  \bibinfo{pages}{093603} (\bibinfo{year}{2008}).

\bibitem[{\citenamefont{Piro et~al.}(2011)\citenamefont{Piro, Rohde, Schuck,
  Almendros, Huwer, Ghosh, Haase, Hennrich, Dubin, and Eschner}}]{piro2011}
\bibinfo{author}{\bibfnamefont{N.}~\bibnamefont{Piro}},
  \bibinfo{author}{\bibfnamefont{F.}~\bibnamefont{Rohde}},
  \bibinfo{author}{\bibfnamefont{C.}~\bibnamefont{Schuck}},
  \bibinfo{author}{\bibfnamefont{M.}~\bibnamefont{Almendros}},
  \bibinfo{author}{\bibfnamefont{J.}~\bibnamefont{Huwer}},
  \bibinfo{author}{\bibfnamefont{J.}~\bibnamefont{Ghosh}},
  \bibinfo{author}{\bibfnamefont{A.}~\bibnamefont{Haase}},
  \bibinfo{author}{\bibfnamefont{M.}~\bibnamefont{Hennrich}},
  \bibinfo{author}{\bibfnamefont{F.}~\bibnamefont{Dubin}}, \bibnamefont{and}
  \bibinfo{author}{\bibfnamefont{J.}~\bibnamefont{Eschner}},
  \bibinfo{journal}{Nat. Phys.} \textbf{\bibinfo{volume}{7}},
  \bibinfo{pages}{17} (\bibinfo{year}{2011}).

\bibitem[{\citenamefont{Stadler et~al.}(2008)\citenamefont{Stadler, Stanciu,
  Stupperich, and Meixner}}]{stadler2008}
\bibinfo{author}{\bibfnamefont{J.}~\bibnamefont{Stadler}},
  \bibinfo{author}{\bibfnamefont{C.}~\bibnamefont{Stanciu}},
  \bibinfo{author}{\bibfnamefont{C.}~\bibnamefont{Stupperich}},
  \bibnamefont{and} \bibinfo{author}{\bibfnamefont{A.~J.}
  \bibnamefont{Meixner}}, \bibinfo{journal}{Opt. Lett.}
  \textbf{\bibinfo{volume}{33}}, \bibinfo{pages}{681} (\bibinfo{year}{2008}).

\bibitem[{\citenamefont{Bokor and Davidson}(2008)}]{bokor2008}
\bibinfo{author}{\bibfnamefont{N.}~\bibnamefont{Bokor}} \bibnamefont{and}
  \bibinfo{author}{\bibfnamefont{N.}~\bibnamefont{Davidson}},
  \bibinfo{journal}{Opt. Commun.} \textbf{\bibinfo{volume}{281}},
  \bibinfo{pages}{5499} (\bibinfo{year}{2008}).

\bibitem[{\citenamefont{Bokor and Davidson}(2004)}]{bokor2004}
\bibinfo{author}{\bibfnamefont{N.}~\bibnamefont{Bokor}} \bibnamefont{and}
  \bibinfo{author}{\bibfnamefont{N.}~\bibnamefont{Davidson}},
  \bibinfo{journal}{Opt. Lett.} \textbf{\bibinfo{volume}{29}},
  \bibinfo{pages}{1968} (\bibinfo{year}{2004}).

\bibitem[{\citenamefont{Urbach and Pereira}(2008)}]{urbach2008}
\bibinfo{author}{\bibfnamefont{H.~P.} \bibnamefont{Urbach}} \bibnamefont{and}
  \bibinfo{author}{\bibfnamefont{S.~F.} \bibnamefont{Pereira}},
  \bibinfo{journal}{Physical Review Letters} \textbf{\bibinfo{volume}{100}},
  \bibinfo{pages}{123904} (\bibinfo{year}{2008}).

\bibitem[{\citenamefont{Basset}(1986)}]{basset1986}
\bibinfo{author}{\bibfnamefont{I.~M.} \bibnamefont{Basset}},
  \bibinfo{journal}{Journal of Modern Optics} \textbf{\bibinfo{volume}{33}},
  \bibinfo{pages}{279} (\bibinfo{year}{1986}).

\bibitem[{\citenamefont{Stamnes and Dhayalan}(1996)}]{stamnes1996}
\bibinfo{author}{\bibfnamefont{J.~J.} \bibnamefont{Stamnes}} \bibnamefont{and}
  \bibinfo{author}{\bibfnamefont{V.}~\bibnamefont{Dhayalan}},
  \bibinfo{journal}{Pure Appl. Opt.} \textbf{\bibinfo{volume}{5}},
  \bibinfo{pages}{195} (\bibinfo{year}{1996}).

\bibitem[{\citenamefont{Sheppard and
  T\"or\"ok}(1997{\natexlab{a}})}]{sheppard1997}
\bibinfo{author}{\bibfnamefont{C.~J.~R.} \bibnamefont{Sheppard}}
  \bibnamefont{and}
  \bibinfo{author}{\bibfnamefont{P.}~\bibnamefont{T\"or\"ok}},
  \bibinfo{journal}{Optik} \textbf{\bibinfo{volume}{104}}, \bibinfo{pages}{175}
  (\bibinfo{year}{1997}{\natexlab{a}}).

\bibitem[{\citenamefont{Quabis et~al.}(2000)\citenamefont{Quabis, Dorn,
  Eberler, Gl\"ockl, and Leuchs}}]{quabis2000}
\bibinfo{author}{\bibfnamefont{S.}~\bibnamefont{Quabis}},
  \bibinfo{author}{\bibfnamefont{R.}~\bibnamefont{Dorn}},
  \bibinfo{author}{\bibfnamefont{M.}~\bibnamefont{Eberler}},
  \bibinfo{author}{\bibfnamefont{O.}~\bibnamefont{Gl\"ockl}}, \bibnamefont{and}
  \bibinfo{author}{\bibfnamefont{G.}~\bibnamefont{Leuchs}},
  \bibinfo{journal}{Opt. Comm.} \textbf{\bibinfo{volume}{179}},
  \bibinfo{pages}{1} (\bibinfo{year}{2000}).

\bibitem[{\citenamefont{van Enk}(2004)}]{vanenk2004}
\bibinfo{author}{\bibfnamefont{S.~J.} \bibnamefont{van Enk}},
  \bibinfo{journal}{Phys. Rev. A} \textbf{\bibinfo{volume}{69}},
  \bibinfo{pages}{043813} (\bibinfo{year}{2004}).

\bibitem[{\citenamefont{Urbach and Pereira}(2009)}]{urbach2009}
\bibinfo{author}{\bibfnamefont{H.~P.} \bibnamefont{Urbach}} \bibnamefont{and}
  \bibinfo{author}{\bibfnamefont{S.~F.} \bibnamefont{Pereira}},
  \bibinfo{journal}{Phys. Rev. A} \textbf{\bibinfo{volume}{79}},
  \bibinfo{pages}{013825} (\bibinfo{year}{2009}).

\bibitem[{\citenamefont{Chen and Zhan}(2010)}]{chen2010b}
\bibinfo{author}{\bibfnamefont{W.}~\bibnamefont{Chen}} \bibnamefont{and}
  \bibinfo{author}{\bibfnamefont{Q.}~\bibnamefont{Zhan}},
  \bibinfo{journal}{Journal of Optics} \textbf{\bibinfo{volume}{12}},
  \bibinfo{pages}{045707} (\bibinfo{year}{2010}).

\bibitem[{\citenamefont{Lerosey et~al.}(2004)\citenamefont{Lerosey, de~Rosny,
  Tourin, Derode, Montaldo, and Fink}}]{lerosey2004}
\bibinfo{author}{\bibfnamefont{G.}~\bibnamefont{Lerosey}},
  \bibinfo{author}{\bibfnamefont{J.}~\bibnamefont{de~Rosny}},
  \bibinfo{author}{\bibfnamefont{A.}~\bibnamefont{Tourin}},
  \bibinfo{author}{\bibfnamefont{A.}~\bibnamefont{Derode}},
  \bibinfo{author}{\bibfnamefont{G.}~\bibnamefont{Montaldo}}, \bibnamefont{and}
  \bibinfo{author}{\bibfnamefont{M.}~\bibnamefont{Fink}},
  \bibinfo{journal}{Phys. Rev. Lett.} \textbf{\bibinfo{volume}{92}},
  \bibinfo{pages}{193904} (\bibinfo{year}{2004}).

\bibitem[{\citenamefont{de~Rosny and Fink}(2002)}]{derosny2002}
\bibinfo{author}{\bibfnamefont{J.}~\bibnamefont{de~Rosny}} \bibnamefont{and}
  \bibinfo{author}{\bibfnamefont{M.}~\bibnamefont{Fink}},
  \bibinfo{journal}{Phys. Rev. Lett.} \textbf{\bibinfo{volume}{89}},
  \bibinfo{pages}{124301} (\bibinfo{year}{2002}).

\bibitem[{Note1()}]{Note1}
  \bibinfo{note}{We note that recently the approach we follow here has
  been applied for the case of an aplanatic lens \cite {chen2010b}.}

\bibitem[{\citenamefont{Sheppard and
  T\"or\"ok}(1997{\natexlab{b}})}]{sheppard1997-a}
\bibinfo{author}{\bibfnamefont{C.~J.~R.} \bibnamefont{Sheppard}}
  \bibnamefont{and}
  \bibinfo{author}{\bibfnamefont{P.}~\bibnamefont{T\"or\"ok}},
  \bibinfo{journal}{J. Mod. Opt.} \textbf{\bibinfo{volume}{44}},
  \bibinfo{pages}{803} (\bibinfo{year}{1997}{\natexlab{b}}).

\bibitem[{\citenamefont{Dhayalan and Stamnes}(1997)}]{dhayalan1997}
\bibinfo{author}{\bibfnamefont{V.}~\bibnamefont{Dhayalan}} \bibnamefont{and}
  \bibinfo{author}{\bibfnamefont{J.~J.} \bibnamefont{Stamnes}},
  \bibinfo{journal}{Pure Appl. Opt.} \textbf{\bibinfo{volume}{6}},
  \bibinfo{pages}{347} (\bibinfo{year}{1997}).

\bibitem[{\citenamefont{Jackson}(1999)}]{jackson1999}
\bibinfo{author}{\bibfnamefont{J.~D.} \bibnamefont{Jackson}},
  \emph{\bibinfo{title}{Classical Electrodynamics}} (\bibinfo{publisher}{Wiley,
  New York}, \bibinfo{year}{1999}), \bibinfo{edition}{3rd} ed.

\bibitem[{\citenamefont{Lieb and Meixner}(2001)}]{lieb2001}
\bibinfo{author}{\bibfnamefont{M.}~\bibnamefont{Lieb}} \bibnamefont{and}
  \bibinfo{author}{\bibfnamefont{A.}~\bibnamefont{Meixner}},
  \bibinfo{journal}{Opt. Express} \textbf{\bibinfo{volume}{8}},
  \bibinfo{pages}{458} (\bibinfo{year}{2001}).

\bibitem[{\citenamefont{Wangsness}(1986)}]{wangsness1986}
\bibinfo{author}{\bibfnamefont{R.~K.} \bibnamefont{Wangsness}},
  \emph{\bibinfo{title}{Electromagnetic fields}} (\bibinfo{publisher}{Wiley},
  \bibinfo{year}{1986}), \bibinfo{edition}{2nd} ed.

\bibitem[{\citenamefont{Leuchs et~al.}(2008)\citenamefont{Leuchs, Mantel,
  Berger, Konermann, Sondermann, Peschel, Lindlein, and Schwider}}]{leuchs2008}
\bibinfo{author}{\bibfnamefont{G.}~\bibnamefont{Leuchs}},
  \bibinfo{author}{\bibfnamefont{K.}~\bibnamefont{Mantel}},
  \bibinfo{author}{\bibfnamefont{A.}~\bibnamefont{Berger}},
  \bibinfo{author}{\bibfnamefont{H.}~\bibnamefont{Konermann}},
  \bibinfo{author}{\bibfnamefont{M.}~\bibnamefont{Sondermann}},
  \bibinfo{author}{\bibfnamefont{U.}~\bibnamefont{Peschel}},
  \bibinfo{author}{\bibfnamefont{N.}~\bibnamefont{Lindlein}}, \bibnamefont{and}
  \bibinfo{author}{\bibfnamefont{J.}~\bibnamefont{Schwider}},
  \bibinfo{journal}{Applied Optics} \textbf{\bibinfo{volume}{47}},
  \bibinfo{pages}{5570} (\bibinfo{year}{2008}).

\bibitem[{\citenamefont{Streed et~al.}(2011)\citenamefont{Streed, Norton,
  Jechow, Weinhold, and Kielpinski}}]{streed2011}
\bibinfo{author}{\bibfnamefont{E.~W.} \bibnamefont{Streed}},
  \bibinfo{author}{\bibfnamefont{B.~G.} \bibnamefont{Norton}},
  \bibinfo{author}{\bibfnamefont{A.}~\bibnamefont{Jechow}},
  \bibinfo{author}{\bibfnamefont{T.~J.} \bibnamefont{Weinhold}},
  \bibnamefont{and}
  \bibinfo{author}{\bibfnamefont{D.}~\bibnamefont{Kielpinski}},
  \bibinfo{journal}{Phys. Rev. Lett.} \textbf{\bibinfo{volume}{106}},
  \bibinfo{pages}{010502} (\bibinfo{year}{2011}).

\bibitem[{\citenamefont{Mandel and Wolf}(1995)}]{mandel-wolf1995}
\bibinfo{author}{\bibfnamefont{L.}~\bibnamefont{Mandel}} \bibnamefont{and}
  \bibinfo{author}{\bibfnamefont{E.}~\bibnamefont{Wolf}},
  \emph{\bibinfo{title}{Optical Coherence and Quantum Optics}}
  (\bibinfo{publisher}{Cambridge University Press},
  \bibinfo{address}{Cambridge, New York}, \bibinfo{year}{1995}), ISBN
  \bibinfo{isbn}{0 521 41711 2}.

\bibitem[{\citenamefont{Hecht}(1987)}]{hecht1987}
\bibinfo{author}{\bibfnamefont{E.}~\bibnamefont{Hecht}},
  \emph{\bibinfo{title}{Optics}} (\bibinfo{publisher}{Addison-Wesley Publishing
  Company}, \bibinfo{year}{1987}).

\bibitem[{\citenamefont{van Enk and Kimble}(2001)}]{vanenk2001}
\bibinfo{author}{\bibfnamefont{S.~J.} \bibnamefont{van Enk}} \bibnamefont{and}
  \bibinfo{author}{\bibfnamefont{H.~J.} \bibnamefont{Kimble}},
  \bibinfo{journal}{Phys. Rev. A} \textbf{\bibinfo{volume}{63}},
  \bibinfo{pages}{023809} (\bibinfo{year}{2001}).

\end{thebibliography}
\end{document}